\documentclass{article}
\usepackage[english]{babel}
\usepackage{float}
\usepackage{tabularx}
\usepackage{booktabs}
\usepackage[a4paper,top=2cm,bottom=2cm,left=3cm,right=3cm,marginparwidth=1.75cm]{geometry}
\usepackage{times}
\usepackage{soul}
\usepackage{color}
\usepackage{graphicx}
\usepackage{ucs}
\usepackage[utf8x]{inputenc}
\usepackage{amsmath}
\usepackage{amstext}
\usepackage{amssymb}
\usepackage{slashed}
\usepackage{mathtools}
\usepackage{enumerate}
\usepackage[normalem]{ulem}
\usepackage{makeidx}
\usepackage[breaklinks=true,colorlinks=false,hidelinks]{hyperref}
\usepackage[english]{cleveref}
\usepackage[font=footnotesize]{caption}
\usepackage{subcaption}
\usepackage{array}
\usepackage[dvipsnames]{xcolor}
\usepackage{dirtytalk}
\usepackage{setspace}
\usepackage{cite}
\usepackage{amsmath}
\usepackage{graphicx}
\spacing{1.5}

\title{\textbf{Revealing giant exciton fine-structure splitting in 2D perovskites using van der Waals passivation}}

\author
{Rodolfo Canet-Albiach$^{1, \dagger}$, Marie Krecmarova$^{1, \dagger}$, José Bosch Bailach$^{1, \dagger}$ \\ Andrés F. Gualdrón-Reyes$^{2, 3}$, Jesús Rodríguez-Romero$^{2, 4}$, Setatira Gorji$^{1}$\\Hamid Pashaei-Adl $^{1}$,  Iván Mora-Seró $^{2}$, Juan Martínez Pastor $^{1,5}$\\ Juan Francisco Sánchez-Royo $^{1,5}$ and Guillermo Muñoz Matutano $^{1,*}$
\\
\normalsize{$^{1}$Instituto de Ciencia de Materiales,
Universidad de Valencia (ICMUV), 46071 Valencia, Spain}\\
\normalsize{$^{2}$Institute of Advanced Materials (INAM), Universitat Jaume I, Avenida de Vicent Sos Baynat,}\\
\normalsize{s/n, 12071, Castelló de la Plana, Spain}\\
\normalsize{$^{3}$Facultad de Ciencias, Instituto de Ciencias Químicas, Isla Teja, Universidad Austral de Chile,}\\
\normalsize{5090000, Valdivia, Chile}\\
\normalsize{$^{4}$Facultad de Química, Universidad Nacional Autónoma de México, Circuito Exterior s/n, C.U.,}\\
\normalsize{Coyoacán, 04510 Mexico City, Mexico}\\
\normalsize{$^{5}$MATINÉE: CSIC Associated Unit (ICMM-ICMUV), Universidad de Valencia, Valencia, Spain}\\
\\
\normalsize{$^\ast$Corresponding author; E-mail:  guillermo.munoz@uv.es}\\
\normalsize{$\dagger$ These authors made equal contributions.}
}

\begin{document}
\maketitle

\begin{abstract}

The study of two-dimensional (2D) van der Waals materials has been an active field of research in the development of new optoelectronics and photonic applications over the last decade. Organic-inorganic layered perovskites are currently some of the most promising 2D van der Waals materials, due to their exceptional optical brightness and enhanced excitonic effects. However, low crystal quality and spectral diffusion usually broaden the exciton linewidth, obscuring the fine structure of the exciton in conventional photoluminescence experiments. Here, we propose a mechanical approach for reducing the effect of spectral diffusion by means of hBN-capping on layered perovskites with different thicknesses, revealing the exciton fine structure. We used a stochastic model to link the reduction of the spectral linewidth with the population of active charge fluctuation centres present in the organic spacer taking part in the dynamical Stark shift. Active fluctuation centres are reduced by a factor of 3.7 to 7.1 when we include hBN-capping according to our direct spectral measurements. This rate is in good agreement with the analysis of the overlap between the squared perovskite lattice and the hexagonal hBN lattice. Van der Waals forces between both lattices cause the partial clamping of the perovskite organic spacer molecules, and hence, the amplitude of the dynamical Stark shift characteristic of the spectral diffusion effect is reduced. Our work provides an easy and low-cost solution to the problem of accessing important fine-structure excitonic state information, along with an explanation of the important carrier dynamics present in the organic spacer that affect the quality of the optical emission.

\end{abstract}

2D van der Waals nanomaterials have attracted considerable attention due to their outstanding electrical and optical properties \cite{mak_photonics_2016}. Layered perovskites belonging to the 2D van der Waals family of materials are the subject of intense scientific research due to their potential to be integrated into high-efficiency photovoltaics and photonic devices \cite{vinattieri_halide_2021}, such as light-emitting diodes (LEDs), lasers or photodetectors \cite{su_perovskite_2021}. Furthermore, 2D layered perovskites can be stacked to form heterostructures with other 2D materials such as transition metal dichalcogenides (TMDCs) \cite{yang_giant_2019,fang_high-performance_2019,andrespenares2021}, graphene \cite{fang_high-performance_2019, solis_de_la_fuente_enhanced_2020} or hexagonal boron nitride (hBN) \cite{leng_molecularly_2018, li_fabrication_2018}, which is among the most used of these types of materials. This engineering strategy provides new routes to design and build novel miniaturised nano-devices with enhanced optoelectronic properties \cite{yang_giant_2019, fang_high-performance_2019, solis_de_la_fuente_enhanced_2020}, strongly stimulating the emergence of new and disruptive applications and scientific breakthroughs. 

Coulomb interaction mediated by excitons present in perovskites is the principal physical mechanism used to generate these enhanced optical emission properties. Intense excitonic interaction should be included to provide an accurate description of the high non-linearity present in both 3D thin films and layered 2D perovskites \cite{ohara_excitonic_2019}. Moreover, the stability of the exciton in 2D perovskites, with binding energies ranging up to several hundreds of meV \cite{ricciardulli_emerging_2021}, makes it possible to study the strong coupling regime of light-matter interaction \cite{su_perovskite_2021} and to develop interesting new states of light and matter at high temperatures, as in, for example, room temperature polariton condensation \cite{su_room-temperature_2017, polimeno_observation_2020}. In a similar direction, electron-hole exchange interaction produces giant exciton fine-structure splitting (FSS), which is measurable even in bulk perovskites \cite{baranowski_giant_2019}. All these features are solid experimental proofs of the enhanced excitonic stability characteristic of perovskite materials.

\begin{figure}
    \centering
    \includegraphics[scale = 0.15]{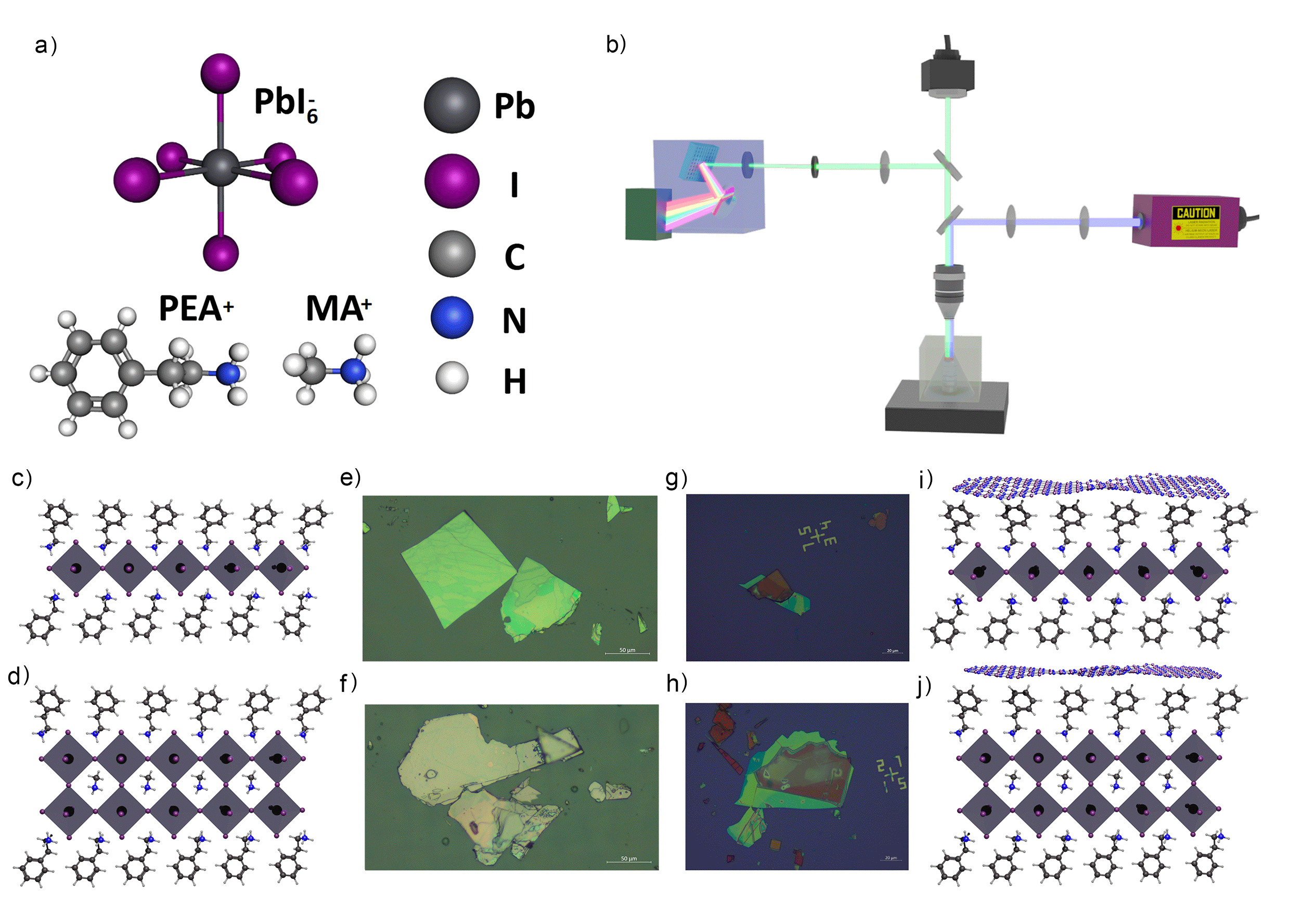}
    \caption{a) Schematic representation of the band structure of 2D halide perovskite, with its excitonic transitions present in the material. Active fluctuation centres are represented as trap states in the organic spacers, later used for our model of the structure. b) Structure of the relevant compounds forming the perovskite structure. 2D perovskites are composed of inorganic octahedral cells (PbI$_6$)$^-$, a short organic methylammonium (MA)$^+$ cation for n $>$ 1 and a long organic phenethylammonium (PEA$_2$)$^+$ cation, which are used as the spacers between the inorganic cells. c) Schematic of the micro-photoluminescence confocal setup used in the experiments. Measurements were performed using a continuous-wave diode laser operated at 405 nm with the sample inside of a closed-cycle helium cryostat. d, e, f, g) 3D representation of the uncapped 2D perovskite (n = 1, 2, respectively) and optical images of representative samples over PDMS substrates. h, i, j, k) 3D representation of the hBN-capped 2D perovskite (n = 1, 2, respectively) and optical images of representative samples over $SiO_2/Si$ substrates. }
    \label{Fig1}
\end{figure}

The particular excitonic energy-level structure of 2D layered perosvkites is the subject of intense and interesting academic debate \cite{baranowski_excitons_2020}. As possible origins of large FSS, both the Rashba effect and electron-hole exchange Coulomb interaction have been considered, modulated by the effective confinement of the quantum well potential created by the 2D layered structure and the dielectric screening \cite{baranowski_excitons_2020}. The S-orbital nature of the minimum conduction band and the valence band, along with the break of the degeneracy of the exciton level by the electron and hole short- and long-range coulombic exchange terms, provides an explanation of the origin of the single, double and triple optical transitions found in the luminescence spectra. These levels are energetically separated by their characteristic FSS, which is directly related to the symmetry properties of the crystal and to the specific optical selection rules. In lower-symmetry perovskite phases, excitonic state degeneracy is broken, producing dark and bright excitonic levels, and state transitions with particular linearly or circularly polarised emissions. Giant excitonic FSS of up to 2 meV has been recently measured in 2D perovskite samples \cite{do_bright_2020}. In that work, the excitonic level structure is modelled and simulated by returning four different bright excitonic transitions (X1, X2, X3 and X4 in Fig. \ref{Fig1}.a) \cite{do_bright_2020}, where both the X2 and X3 states are energetically split into two linearly polarised optical transitions. This extra splitting has been associated with exciton FSS. However, the appearance of these extra peaks measured in the photoluminescence spectra can be associated with alternative physical mechanisms. The softness of the ionic perovskite lattice gives rise to strong exciton-phonon coupling, which has been proposed to describe dressed exciton-polaron formation in two-dimensional hybrid perovskites \cite{thouin_phonon_2019}. This mechanism leads to vibronic resonances associated with the excitonic coupling with low-frequency optical phonons. Also, it has been shown avoided crossing between high-order natural cavity modes and excitonic transitions in the reflectivity and luminescence spectra, as a function of 2D perovskite thickness \cite{anantharaman_self-hybridized_2021}. This is the principal signature of the formation of low and upper polaritons as the dressed basis describing strong coupling between light modes and excitonic states. As a consequence, deeper studies and debate are needed to provide a clearer picture of the physical processes enclosed in the complex fine structure of the 2D layered perovskite excitonic-level structure. 

In this work, we have measured the micro-photoluminescence ($\mu$-PL) spectra of 2D layered perovskites with different phase thicknesses. We compared these spectra with the spectra produced when hBN is included on top of the perovskites as a secondary isolating 2D structure. hBN-capping affects the photoluminescence spectra, returning four different excitonic transitions. The ability to compare the photoluminescence linewidths of excitonic transitions in hBN-capped and uncapped layered perovskites gives us the ability to analyse two different carrier dynamical situations by spectral means. Other works have also studied the effects of the addition of hBN to heterostructures, focusing on the enhancement of stability due to protection from the environment in 2D halide perovskites \cite{seitz_long-term_2019} or the suppression of broadening effects to enhance the linewidth photoluminescent response in 2D Transition metal dichalcogenide monolayers\cite{hoshi_suppression_2017}, but to our knowledge, none have focused on understanding the mechanism by which this optical quality enhancement works. In addition, we simulated the experimental data with a stochastic model in which carrier fluctuation in the organic layers affects the excitonic energy transition due to the randomised dynamical Stark effect and thus produces spectral diffusion at each one of the excitonic optical transitions. As the number of active fluctuation centres is attenuated by the hBN mechanical passivation of the perovskite surface, the strength of the spectral diffusion is reduced, and then the FSS of the excitonic recombination is revealed. Following this procedure, we have measured a giant FSS of 2 meV, which is compatible with recent reports concerning 2D perovskite single-crystal growth using chemical synthesis with minimal surface fluctuation and low defect levels. By the analysis of the Fourier transform of the first-order correlation function, fitted to our experimental data, we can replicate the change of the peak behaviour from Gaussian to Lorentzian. Using our model, we can associate the measured excitonic linewidth narrowing with the density reduction of the active fluctuation centres by a factor of 3.7 to 7.1 in our different samples. This analysis is in good  agreement with the geometrical evaluation of organic molecule passivation by van der Waals clamping using a simple geometrical model that accounts for lattice overlap between the materials. In summary, our study provides very relevant and solid experimental information to the actual open discussion of the structure of excitonic optical transitions. It is crucial for helping the community to understand the physical phenomena underlying the origin of giant FSS and the possible strategies that can be used to reveal  giant FSS. The control of states with giant exciton FSS will play an important role in the development of spintronics and spin quantum computing using low-dimensional semiconductors\cite{astakhov_exciton_2007}.

\section*{Results and discussion}

\subsection*{Synthesis and sample preparation}

The two-dimensional perovskite $(PEA)_2(MA)_{n-1}Pb_nI_{3n + 1}$ crystals were synthesised following the procedure described on the Methods section. Here, $n$ refers to the number of perovskite layers inside this structure, known as the Ruddlesden-Popper (RP) phase, where perovskite layers are held together by molecular cations. Figure \ref{Fig1}.c shows the structure of a single 2D layer (n = 1), where phenethylamine is being used as the cation (PEA, Fig. \ref{Fig1}.a). In the case of n $\geq$ 2, i.e., 2D structures with an increasing number of perovskite layers, methylamine (MA, Fig. \ref{Fig1}.a) is found between the layers. Figure ~\ref{Fig2}.d shows the structure of a double perovskite layer (n = 2). This structure allows for strong intra-layer forces but weak inter-layer forces, as the long organic chains are stuck to one another by van der Waals forces. Figure \ref{Fig1}.e, f shows optical images corresponding to 2D perovskite samples with phase thicknesses of n = 1 and n = 2. We use hexagonal boron nitride (hBN) as a second 2D material. After localising and deciding which hBN samples were suitable for the experiment, they were transferred from PDMS onto perovskite flakes following the all-dry viscoelastic method \cite{castellanos-gomez_deterministic_2014} in a custom transference setup. Figure \ref{Fig2}.i, j shows a schematic view of the process, where hBN is represented as a monolayer; it is being placed on top of 2D perovskites with n = 1 and n = 2, respectively. The samples were not fully covered with hBN to keep a part of the sample free from hBN-capping. This allowed us to study both capped and uncapped situations from the same sample. Optical images of samples with hBN-capping for samples with phase thicknesses of n = 1 and n = 2 are shown in Fig.~\ref{Fig1}.g, h.

$\mu$-PL measurements were taken in a custom fiber-based confocal setup (Fig.~\ref{Fig1}.b) equipped with a closed-cycle helium cryostat. All $\mu$-PL measurements were done at 4 K. The measurements were performed using a continuous-wave diode laser operated at 405 nm. Excitation and detection were done using a long-working-distance 50$\times$ microscope objective lens with a numerical aperture of NA = 0.42 that was mounted outside the cryostat. The emission from the sample was long-pass filtered, dispersed by a double-grating spectrograph with a focal length of 0.3 m (Acton SP-300i from Princeton Instruments) and detected with a cooled Si CCD camera (Newton EMCCD from ANDOR).

\subsection*{micro-Photoluminescence}

\begin{figure}
    \centering
    \includegraphics[scale = 0.35]{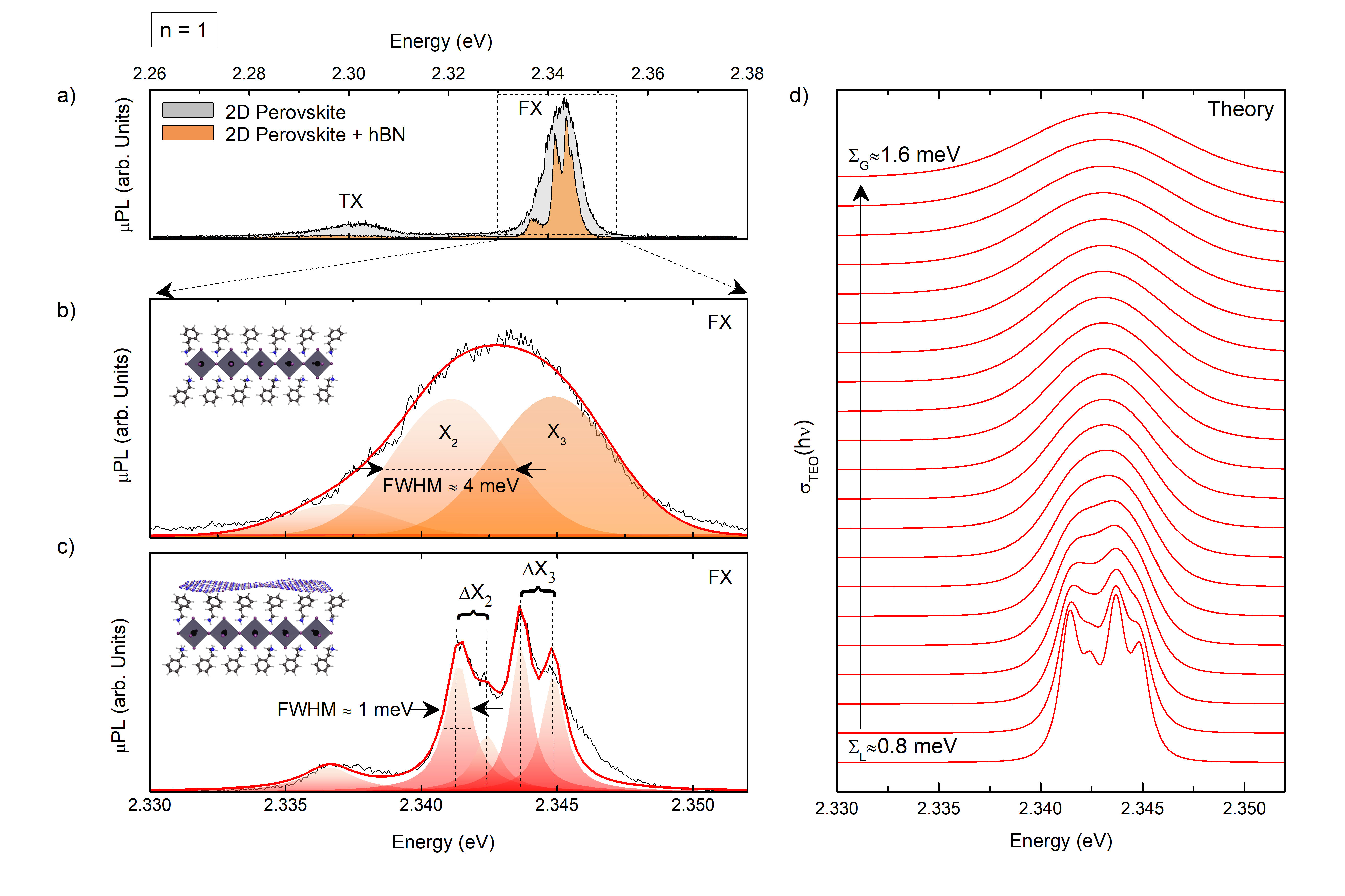}
    \caption{a) Photoluminescence spectra of a representative sample of 2D perovskite (n = 1) for both uncapped (uncoloured spectrum) and hBN-capped (coloured spectrum) areas. It shows two main features, which correspond to trapped excitons and free excitons, respectively. Measurements were performed at 4 K by a 405 nm continuous-wave laser at 1.5 $\mu$W of power over a $SiO_2/Si$ substrate. b) Close-up of the previous photoluminescence spectra for uncapped perovskite (n = 1). c) Close-up of the previous photoluminescence spectra for hBN-capped perovskite (n = 1). d) Modelled transition from Gaussian to Lorentzian behaviour in the n = 1 case, fitted to the uncapped and hBN-capped behaviour with an FWHM $\approx$ 1 meV to 4 meV.}
    \label{Fig2}
\end{figure}

Figure~\ref{Fig2}.a shows the $\mu$-PL spectra from a uncapped perovskite and hBN-capped perovskite for n = 1. Spectra were recorded at 4 K with an excitation power of 1.5 $\mu$W. As can be seen, the uncapped spectrum reveals two main transitions. The first, centred around 2.305 eV, is attributed to trapped excitons (TX), and the second, centred around 2.343 eV, corresponds to free excitons (FX) \cite{gorji_donor-acceptor_nodate}. Do et al. recently showed how spin orbit coupling splits the excitonic transition in 2D halide perovskites into four different recombination paths (X1, X2, X3 and X4) \cite{do_bright_2020}. Here, the literature-reported X1 transition is equivalent to our TX transition. The X4 transition should lay on the high-energy side ($\sim$2.38 eV). Finally, the X2 and X3 transitions are compatible with our FX energy range. Here, the spectra could be modelled by the combination of three Gaussian peaks, with the two latter peaks corresponding to the X2 and X3 transitions, both with an FWHM of $\sim$ 4 meV, and the first peak corresponding to phonon contributions \cite{thouin_phonon_2019}. As will be proved below, the emission from these transitions is mixed and hidden by the effects of spectral diffusion. However, once the measurements are done over the hBN-capped area of the perovskite, the FX peak is divided into five different transitions, as shown in the last graph in Fig. \ref{Fig2}.a, which can be fitted to a Gaussian, the previously mentioned phonon contributions and four different Lorentzian peaks. The positions of these Lorentzian peaks have been recorded and are shown in Fig. \ref{Fig4}.b, where they have been classified as the transitions corresponding to the upper and lower subbands of $X_2$ and $X_3$ in four different flakes. These transitions present a full width at half maximum (FWHM) of around 1 meV, which corresponds to a four-fold decrease in the FWHM, showing a clear contrast to the wider Gaussian in the uncapped spectrum. These values are in agreement with the recent literature\cite{do_bright_2020}. We have labelled $\Delta X_2$ and $\Delta X_3$ as the energy shifts of each subband. Our statistical analysis is consistent between different flakes, and the average values of these energy shifts are $$\Delta X_2^{m, n=1} = 1.00 \pm 0.10 \; \text{meV},$$ $$\Delta X_3^{m, n = 1} = 1.25 \pm 0.10 \;\text{meV}.$$

\begin{figure}
    \centering
    \includegraphics[scale = 0.35]{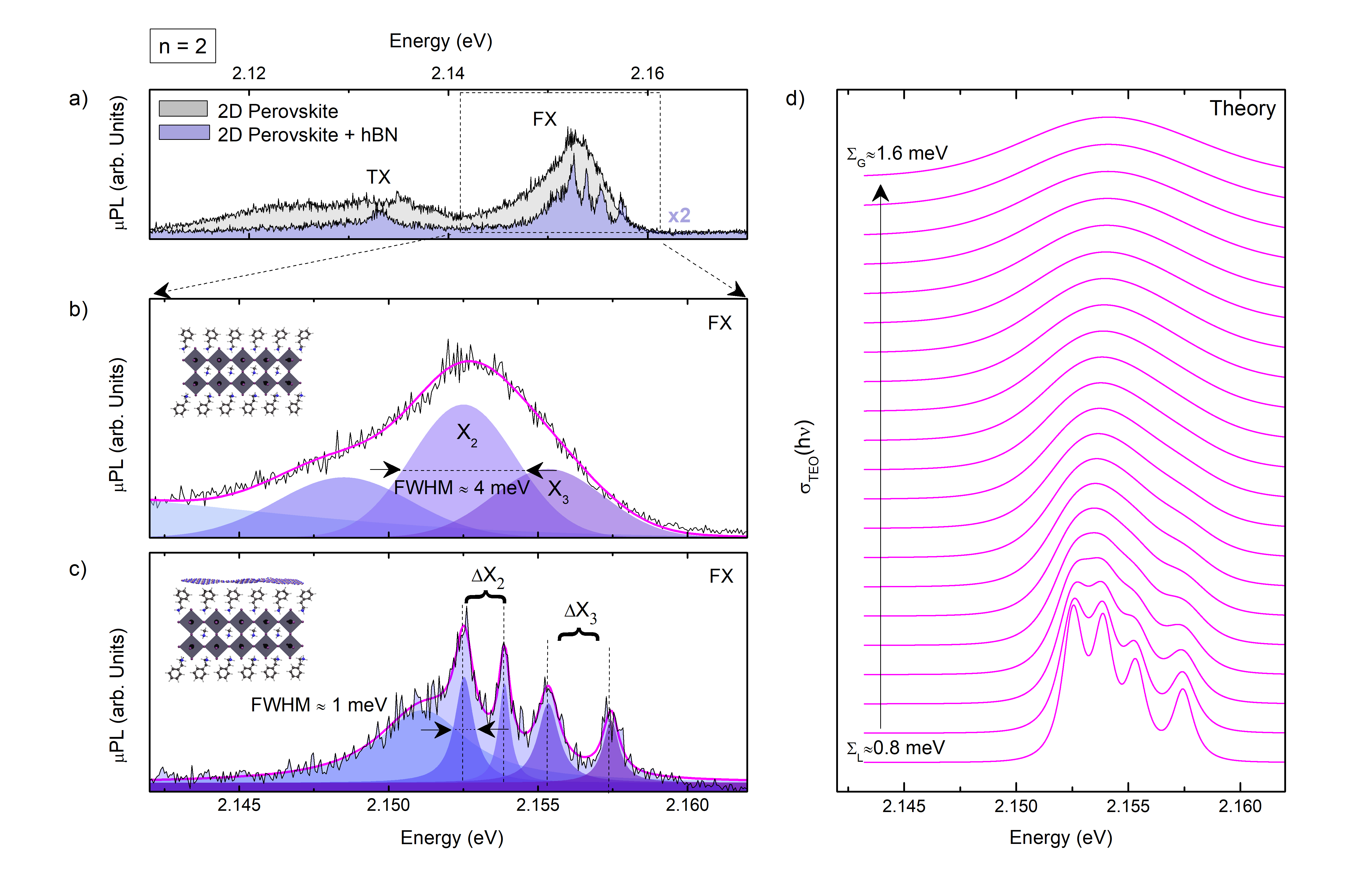}
    \caption{a) Photoluminescence spectra of a representative sample of 2D perovskite (n = 2) for both uncapped (uncoloured spectrum) and hBN-capped (coloured spectrum) areas.  It shows two main features, a broad transition corresponding to trapped excitons and a peak that corresponds to free excitons. Measurements were performed at 4 K by a 405 nm continuous-wave laser at 0.8 $\mu$W of power over a $SiO_2/Si$ substrate. b) Close-up of the previous photoluminescence spectra for uncapped perovskite (n = 2). c) Close-up of the previous photoluminescence spectra for hBN-capped perovskite (n = 2). d) Modelled transition from Gaussian to Lorentzian behaviour in the n = 2 case, fitted to the uncapped and hBN-capped behaviour with an FWHM $\approx$ 1 meV to 4 meV.}
    \label{Fig3}
\end{figure}

To further explore the effect of hBN-capping on 2D perovskites, more measurements were done on RP phase n = 2, as shown in Fig.~\ref{Fig3}.a with uncapped and hBN-capped samples. Spectra have been recorded at 4 K with an excitation power of 800 nW. As in the previous case, the uncapped spectrum shows two main features. Here, TX recombination shows a broader optical transition. It is difficult to assign a clear transition peak, as the TX optical recombination covers an energy band from 2.11 to 2.14 eV. As we discussed previously, thicker phases with $n \geq 2$ are affected by a higher carrier complexity, where the donor-acceptor pair optical transition is activated over the range between FX and TX\cite{gorji_donor-acceptor_nodate}. Here, FX recombination is fitted with a Gaussian peak centred around 2.152 eV. The fitting reveals the combination of three different Gaussians in our FX peak. As in the n = 1 case, the latter two correspond to the X2 and X3 transitions, with an FWHM of $\sim$ 4 meV. In this phase, the X3 transition seems to emit with lower intensity than X2. As in the n = 1 case, the spectra of hBN-capped perovskite reveal five different transitions, as shown in the bottom panel of Fig.~\ref{Fig3}.a. The latter four are again better fitted with Lorentzian functions. As before, we labelled them as the two doublets from each X2 and X3 subband energy shift. In the second half of Fig. \ref{Fig4}.b, the statistics of FSS in our n = 2 flakes are shown. Their values, $\Delta X_2$ and $\Delta X_3$, are in general homogeneous. Now, due to the higher carrier complexity associated with this thickness phase, the Lorentzian FWHM greatly differs for each of the transitions: it is equal to $0.77 \pm 0.09$, $0.46 \pm 0.05$, $0.87 \pm 0.09$ and $0.66 \pm 0.08$ meV. The experimental average FSS values are $$\Delta X_2^{m, n=2} = 1.43 \pm 0.10 \; \text{meV},$$ $$\Delta X_3^{m, n = 2} = 1.56 \pm 0.10 \;\text{meV}.$$ In both the n = 1 and n = 2 cases, these FSS values are comparable to values from other studies on two-dimensional perovskites \cite{do_bright_2020}, but they are larger in magnitude than other FSS values from perovskite nanocrystals \cite{fu_neutral_2017,becker_bright_2018, ramade_fine_2018} and other materials \cite{granados_del_aguila_linearly_2019,htoon_anomalous_2009}.

\subsection*{Stochastic and geometrical model simulation}

To provide an explanation of the origin of the changes in lineshapes and linewidths in 2D perovskites for both n = 1 and n = 2 when we include hBN-capping and hence reveal the excitonic FSS, we compared our measured spectra with the output of a model simulation. We adapted the stochastic model of Berthelot et al.\cite{berthelot_unconventional_2006} to quantitatively evaluate linewidth broadening due to the effects of spectral diffusion (SD). It is well known that SD is a source of optical transition linewidth broadening in semiconductors, like Quantum Dots (QDs) \cite{berthelot_unconventional_2006,holmes_spectral_2015}, and 2D materials \cite{spokoyny_effect_2020, he_phonon_2016}. Dynamical charge fluctuations in the surroundings of the nanostructure build a randomised electric field that produces a stochastic AC Stark effect. The standard procedure to evaluate the origin and magnitude of the SD is time-domain analysis, which produces a study of the lineshape evolution of the temporal correlation function (g(t)) \cite{berthelot_unconventional_2006}. However, our ability to compare two spectral situations from the same flake allows the direct spectral analysis of the SD, and thus we compare our data with the Fourier transform of the temporal correlation function of the model, i.e., with the theoretical absorption spectrum ($\sigma (\omega)$) \cite{kubo_ryogo_stochastic_1962} : 
\begin{gather*}
    g(t) = \frac{\Sigma^2\tau_c^2}{\hbar^2}(\frac{t}{\tau} + e^{-\frac{t}{\tau}}- 1),\\
    \sigma(\omega) = \int^{\infty}_{0}e^{-g(t) + it(\omega - \omega_0)}dt,
\end{gather*}
where $\omega_0$ represents the frequency that corresponds to each excitonic transition, $\Sigma$ represents the spectral amplitude of the transition and $\tau_c$ represents the correlation time, the magnitude of the timescale of the SD processes, which is directly linked with the timescale of the charge changes of the active fluctuation centres. There are two extreme scenarios of the g(t) function (Gaussian for $\tau_c\Sigma >> \hbar$, and an exponential decay for $\tau_c\Sigma << \hbar$): the absorption lineshape becomes Gaussian in the first case and Lorentzian in the second. Here, we use the spectral amplitude $\Sigma$ as the model free parameter; its extreme values were limited by our experimentally measured FWHM. In our case, we suppose a fixed correlation time, and thus assume that the variation of the SD magnitude is only linked with the different active fluctuation centres present in the organic 2D layer. For our model, the input parameters are the frequencies $\omega_i$ corresponding to the energies of the transitions and the FWHMs of both the Gaussian ($FWHM_{G}$) and Lorentzian ($FWHM_{L}$), which are proportional to the spectral amplitudes ($\Sigma_G$ and $\Sigma_L$, respectively). From the limit criteria for Gaussian behaviour, we obtain a lower limit of 0.5 ps for $\tau_c$, which we consider as a constant for both capped and uncapped situations.

\begin{figure}
    \centering
    \includegraphics[scale = 0.45]{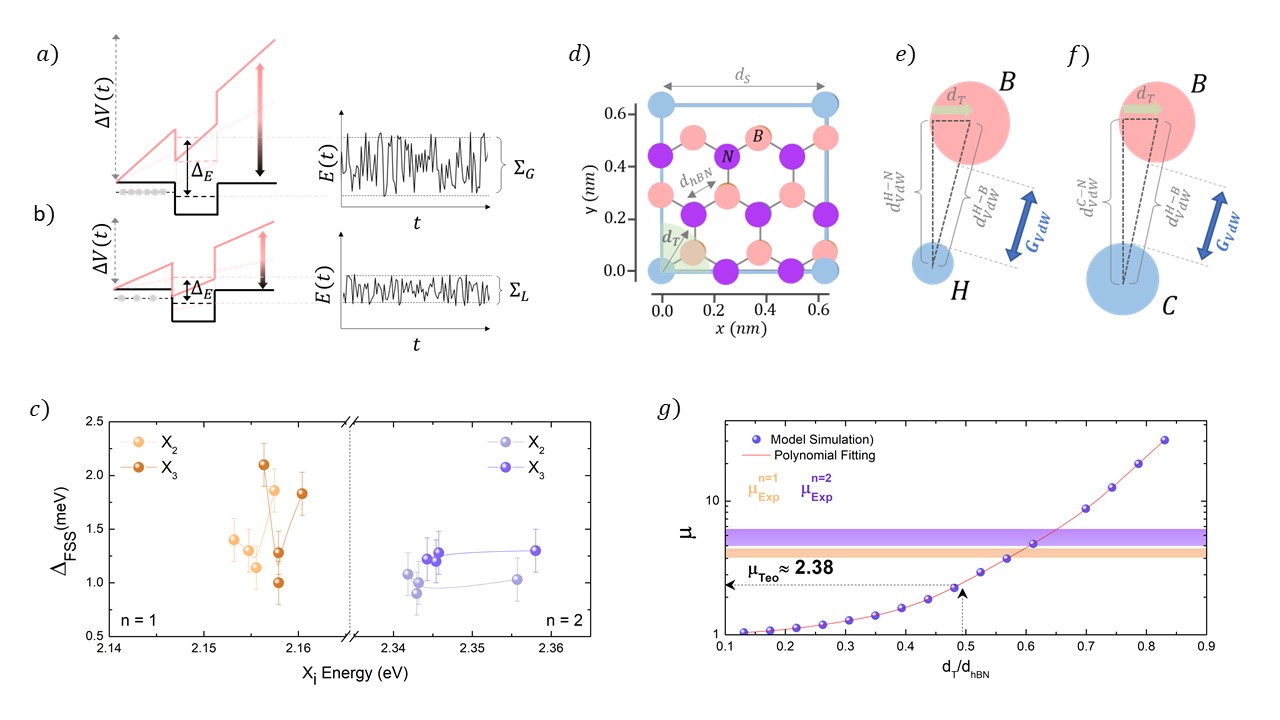}
    \caption{a, b) Schematic representation of the band structure being altered by a randomised AC field created by the Stark effect of the molecular reorientation of the organic spacers. For the Gaussian behaviour (a) the number of active fluctuation centres is larger, and hence the amplitude of the Stark energy shift ($\Delta_{E}$) is larger, too. For the Lorentzian behaviour (b) partial molecular clamp reduces the amplitude of the Stark energy shift. In both figures we have simulated the stochastic evolution of the energy state as a function of time ($E(t)$), where the spectral amplitude of the transition is labelled as $\Sigma_{G/L}$. c) Representation of the experimental values for the fine-structure splitting observed in our set of samples for both n = 1 and n = 2. d) Geometrical matching of the hexagonal lattice from hBN and the square lattice for 2D perovskite spacers ($d_S$ = 0.632 nm, $d_hBN$ = 0.145 nm). e, f) Geometrical calculation of the numerical value for the optimal tolerance distance for the distance between atoms, for both carbon and hydrogen atoms ($d_{vdW}^{H-N} = 0.312$ nm, $d_{vdW}^{H-B} = 0.320$ nm, $d_{vdW}^{C-N} = 0.352$ nm, $d_{vdW}^{C-B} = 0.359$ nm, $d_T = 0.071$ nm). g) Simulation results for the average coincidence ratio versus the ratio of the tolerance distance and the lattice constant from hBN. A polynomial fitting was then applied to find the exact value of the distance ratio, which is 0.489 times the lattice constant for hBN, which is $\sim$0.071 nm.}
    \label{Fig4}
\end{figure}

 Through the dynamical charge fluctuation mechanism, the aforementioned stochastic AC Stark effect is activated (Fig. \ref{Fig4}.a). This results in random changes of the energy levels and thus an effective linewidth broadening of each excitonic transition. If the magnitude of the electric field is high enough, i.e., if the number of active fluctuation centres $N$ is large enough, a situation where both the $X_2$ and $X_3$ transitions can optically overlap may be reached. Considering that the energy shift associated  with a single active fluctuation centre is $\Delta$, a relation between $N$ and the spectral amplitude $\Sigma$ can be obtained. We can estimate the ratio of active centres before ($N_G$) hBN-capping to the ratio of centres after hBN-capping ($N_L$) for both n = 1 and n = 2. Following this procedure, the ratio of active centres in the organic layer is expressed as \cite{berthelot_unconventional_2006} $$\mu_{Exp}=\frac{N_G}{N_L} = \left(\frac{\Sigma_G}{\Sigma_L}\right)^2.$$
 
 With the model fitted to our experimental data, we can recreate the different states of the system for a range of spectral amplitudes, showing the changes between Gaussian and Lorentzian behaviour. As shown in Fig. \ref{Fig2}.b, we can see the modelled transition from four Lorentzian peaks to a single wider Gaussian peak formed by the contributions of the Lorentzian peaks by gradually changing the FWHM of the model from $\sim$1 meV to $\sim$4 meV. As shown in both Fig. \ref{Fig2}.b and Fig. \ref{Fig3}.b, the model output nicely reproduces our measured spectra for both capped and uncapped situations. From our experimental data, $\mu^{n = 1}_{Exp} = (3.7 \pm 0.2); (3.7 \pm 0.4); (4.9 \pm 0.4)$ and $(4.4 \pm 0.4)$ for each of the four transitions in the n = 1 case. In the case of n = 2, $\mu^{n = 2}_{Exp} = (4.4 \pm 0.8); (7.3 \pm 1.2); (3.9 \pm 0.6);$ and $(5.1 \pm 0.9)$. This is a more dispersed set of values than those from the n = 1 case, which is attributable to the higher variability of the FWHM for the Lorentzian peaks in the n = 2 case. This variability of the FWHM could in turn be attributed to the higher complexity of the carrier dynamics for the n = 2 case, due to the addition of methylammonium to the octahedral perovskite structure \cite{gorji_donor-acceptor_nodate}.

Once the model analysis was completed, we investigated the source of the active fluctuation centres in our material. As a possible physical origin, we tentatively associate them with the site-specific dynamics of the motion of the organic spacers. Molecular movement is somewhat restricted thanks to van der Waals forces between the different organic layers. The length of the organic spacer and its relative van der Waals forces, and hence its molecular rigidity, play an important role in the structural and electronic disorder in RP phases \cite{dahlman_dynamic_2021}. For PEA cations, the timescale of these molecular fluctuations is on the order of microseconds\cite{dahlman_dynamic_2021}, which would be slow enough to produce excitonic linewidth broadening by SD, as the excitonic decay time is on the order of hundreds of ps \cite{gorji_donor-acceptor_nodate}. The hBN layers will clamp part of the upper-layer PEA molecules. This produces a more stable spacer environment, reducing the number of active molecules in movement, i.e., reducing the number of active fluctuation centres previously described in our model. Any of the atoms present in hBN (B or N), by interacting with the external atoms in the aromatic ring of the PEA molecule (either C or H, see Fig. \ref{Fig1}.a), could clamp the molecule and hence reduce the amplitude of its movement. 

We performed numerical simulations to estimate the possible magnitude of the ratio of active fluctuation centres (non-clamped molecules) in uncapped perovskite to active fluctuation centres in hBN-capped perovskite. We built a basic model consisting of the spatial overlapping of the hexagonal lattice of hBN  and the square lattice of the iodine atoms. Here, we suppose that PEA molecules are replicating the same squared lattice. We include the possibility of adding an angle $\theta$ between both lattices. In order to estimate the inter-lattice separation, we have calculated van der Waals interatomic distances between B or N in the hexagonal lattice and H or C in the aromatic molecule that are derived from their atomic radii and the van der Waals bonding gap ($G_{vdW}$, assumed to be $\sim$0.2 nm)\cite{liu_van_2019}. From the distinct values, we calculated an inter-lattice separation of 0.335 nm, the average of all of these values.

We have run these numerical simulations to count planar coincidences between H/C atoms in the squared lattice and B/N atoms in the hexagonal lattice (for a complete description of the numerical evaluation, see the Supp. Information). In our simulations, we built lattices large enough to not be affected by Moiré effects. The basic assumptions for the calculation are shown in Fig. \ref{Fig4}.d. We defined the tolerance distance in the xy plane $d_{T}$ as the maximum accepted distance between an atom in the squared lattice and a second atom in the hexagonal lattice that were considered to make up a planar coincidence. To calculate $d_{T}$, we assumed that the layers are separated by the shortest van der Waals interatomic distances. As can be seen in Fig. \ref{Fig4}.e, f, with interatomic distances between either H-N, H-B, C-N or C-B, the resulting optimal tolerance in the xy distance is $d_{T} = 0.49\times d_{hBN} = 0.071$ nm for both of them. We ran the simulation to account for all possible coincidences between atoms in the perovskite layer and the hexagonal layer within $d_{T}$. Finally, we calculated the ratio between the total number of atoms in the free perovskite and the number of atoms that are not fixed by van der Waals forces when the hexagonal lattice is included on top, i.e.,
\begin{gather*}
    \mu_{Theo}=\frac{N_{sq. lat.}}{N_{sq. lat.} - N_{coincidences}}
\end{gather*}
where $N_{sq. lat.}$ represents the total number of atoms in the square lattice. Our model output shows that the parameter $\mu_{Theo}$ does not suffer any dramatic variation when the angle between the lattices is changed (see the Supp. Information). Figure \ref{Fig4}.g shows the numerical evaluation of $\mu_{Theo}$ as a function of the tolerance distance $d_{T}$. An arrow is used to highlight the expected $\mu_{Theo}$ for $d_{T} = 0.49\times d_{hBN}$, which is $\sim$2.38. Reddish and bluish shadows are used to show the range of the experimentally measured $\mu_{Exp}$ calculated above. As shown by the figure, $\mu_{Exp}$ and $\mu_{Theo}$ are in moderate agreement. This result reinforces our assumption that the AC Stark shift is reduced by the clamping of PEA molecules due to van der Waals forces.

\section*{Conclusions}

In summary, the exciton fine structure of $(PEA)_2(MA)_{n-1}Pb_nI_{3n + 1}$ 2D perovskites has been measured, thanks to a novel hBN-capping approach that directly affects the carrier dynamics of organic-inorganic interaction. Other studies of hBN-capped perovskites have focused on protection from degradation and other extrinsic factors \cite{seitz_long-term_2019}. However, our results show an easy, low-cost approach for producing an effective enhancement of the optical quality of the emission in 2D perovskites. We explained our findings using the effect of the reduction of the SD amplitude, which is caused by the molecular motion clamping of the organic layer by means of van der Waals forces. Our results are of great importance both for future applicability and fundamental research. On the one hand, we provide an alternative way to explore giant exciton FSS and thus interesting spin physics phenomena, if low-defect and minimal-surface-fluctuation flakes are not easily achievable. On the other hand, the agreement between our experimental data and the models can be further exploited to perform a more accurate analysis of the organic-inorganic carrier processes affecting optical transitions or to discover the relevance of the intense excitonic and light-matter interaction effects present in this promising 2D material.

\section*{Materials and Methods}

\subsection*{Synthesis}
Materials were prepared by modifying a synthetic procedure reported elsewhere \cite{stoumpos_ruddlesdenpopper_2016}. The following discusses the synthesis of  PEA$_{2}$PbI$_{4}$ (n = 1).  For Solution A, PbO powder (1116 mg, 5 mmol) was dissolved in a mixture of 57\%
w/w aqueous HI solution (5.0 mL, 38 mmol) and 50$\%$ aqueous H$_3$PO$_2$ (0.85 mL, 7.75 mmol) by heating it to boiling under vigorous stirring for about 10 min. Then, it is possible to see the formation of a bright yellow solution. For Solution B, in a separate beaker, C$_{6}$H$_{5}$CH$_{2}$CH$_{2}$NH$_{2}$ (phenetylamine, PEA) (628 $\mu$L, 5 mmol) was neutralised with HI 57\% w/w (2.5 mL, 19 mmol) in an ice bath, resulting in a clear pale-yellow solution. If  a solid precipitate / the formation of a suspension is observed, it is possible to heat it slightly until it dissolves. Once both solutions have been prepared, we proceed to slowly add Solution B to Solution A. The quick addition of Solution A could produce a precipitate, which could be subsequently dissolved by heating the combined solution to boiling. After this, the solution was kept at the boiling point for 10 min;  stirring was then discontinued, and the solution was left to cool to room temperature. During this time, orange rectangular-shaped plates started to crystallise. The crystallisation was deemed to be complete after 1 hour. The crystals were isolated by suction filtration and thoroughly dried under reduced pressure.  The obtained single crystals were washed with cold diethyl ether.

The following concerns the synthesis of PEA$_{2}$MAPb$_{2}$I$_{7}$ (n = 2).  For Solution A, PbO powder (1116 mg, 5 mmol) was dissolved in a mixture of 57\% w/w aqueous HI solution (5.0 mL, 38 mmol)  and 50\% aqueous H$_{3}$PO$_{2}$ (0.85 mL, 7.75 mmol) by heating it to boiling under vigorous stirring for about 5 min and forming a yellow solution. Then, CH$_{3}$NH$_{3}$Cl powder (169 mg, 2.5 mmol) was added very slowly to the hot yellow solution, causing the precipitation of a black precipitate, which rapidly redissolved under stirring to create a clear, bright yellow solution. For Solution B, in a separate beaker, C$_6$H$_5$CH$_2$CH$_2$NH$_2$ (PEA) (880 $\mu$L, 7 mmol) was neutralised with HI 57 \% w/w (2.5 mL, 19 mmol)  in an ice bath, resulting in a clear pale-yellow solution. The addition of the PEAI solution to the PbI$_{2}$ solution initially produced a black precipitate, which was dissolved by heating the combined solution to boiling. After discontinuing stirring, the solution was cooled to room temperature and cherry-red crystals started to emerge. The precipitation was deemed to be complete after $\sim$1 hour. The crystals were isolated by suction filtration and thoroughly dried under reduced pressure. The material was washed with cold diethyl ether.

\subsection*{Crystal cleaving and sample preparation}
2D perovskite crystals ($PEA_{2}PbI_{4}$ and $PEA_{2}MAPb_{4}I_{7}$) were mechanically exfoliated on Si substrates using Nitto Tape, a type of scotch tape.

hBN was purchased from HQ Graphene, and it shows similar weak inter-layer forces. Thanks to these weak forces, both hBN and the perovskite flakes are easily exfoliable. We used the well-known scotch tape technique. After the initial exfoliation, a different transfer strategy was used in each case. The two-dimensional perovskite flakes were transferred onto a clean $SiO_2/Si$ substrate; this is the usual substrate used for other two-dimensional materials, like transition metal dichalcogenides (TMDs) or graphene. hBN was transferred onto a layer of polydimethylsiloxane (PDMS) to prepare it for transference to the corresponding flake of two-dimensional perovskite. For localisation, a Zeiss Axio Scope.A1 microscope with an Axiocam Erc 5s camera was used. Samples were then selected according to their lateral size and smoothness, as their photoluminescent response does not change depending on the number of layers present. The typical measured thickness was around tens of micrometers.

\subsection*{Optical spectroscopy \& confocal optical microscopy}
All low-temperature measurements were performed with a conventional micro-PL setup, where the samples were held in the cold finger of a vibration-free closed-cycle cryostat (AttoDRY800 from Attocube AG)\cite{pashaei_adl_homogeneous_2021}. The samples were excited by means of a continuous wave (pulsed) excitation laser at a wavelength of $\lambda$ = 405 nm (450 nm), resulting in spectral (time-resolved) PL measurements. Moreover, the alignment of the collection and excitation spots was done for 2D perovskite (n = 1, 2) using single-mode fiber-coupled continuous wave lasers at a wavelength of $\lambda$ = 532 nm. The single-mode optical fibers act as the confocal pinholes during excitation and detection. Excitation and detection were done using a long-working-distance 50$\times$ microscope objective lens with a numerical aperture of NA = 0.5 that was mounted outside the cryostat. The emission from the sample was long-pass filtered, dispersed by a double-grating spectrograph with a focal length of 0.3 m (Acton SP-300i from Princeton Instruments) and detected with a cooled Si CCD camera (Newton EMCCD from ANDOR) to record PL spectral data. A SPAD detector (from Micro Photon Devices) connected to a time-correlated single-photon-counting electronic board (TCC900 from Edinburgh Instruments) was used to obtain time-resolved PL measurements.

\subsection*{Model simulation}

By using the Python package `pybinding', we set up a model heterostructure formed by a monolayer of hBN, with its hexagonal lattice, on top of a monolayer of non-specified atoms symbolising the top layer of organic spacers, which have a square lattice, as we assume that the organic spacers involved will approximately follow the same lattice as the perovskite octahedra to which they are attached. The layer of spacers is twisted with respect to the hBN layer by a variable angle $\theta$. 

The structural position of every atom in each lattice was used along with a distance calculation algorithm  to account for planar coincidences between the positions of atoms in the plane XY. The simulation size, twisting angle and tolerance distances were varied to obtain diverse sets of $\mu$ for each set of parameters. The twisting angle was found to be of little importance at the largest simulation size (30 x 30 nm, 45 x 45 nm), so the average of the angles was calculated and used.

\section*{Acknowledgments}

This work was made possible by the project 2D-SPD (Two-Dimensional Semiconductor Photonic Dots), which was funded by the Spanish Ministry of Science (MICINN), Agencia Estatal de Investigación (AEI) (RTI2018-099015-J-I00) and the European Regional Development Fund (ERDF); the Horizon 2020 Research and Innovation Program of the EU through the S2QUIP project (grant agreement No. 8204023); the Ministerio de Ciencia e Innovación, which is part of AEI, through the project PID2020-112507GB-I00 (Novel quantum states in heterostructures of 2D materials)and PID2019-
107314RB-I00 (Stable); and the Generalitat Valenciana through the project PROMETEU/2021/082. M. K. acknowledges the APOSTD programme (contract APOSTD/2020/103).

\bibliographystyle{naturemag}
\bibliography{Bibtex}

\section*{Supplemental Information}

\subsection*{Model information}

The shift in the dynamical behaviour of the hBN / 2D perovskite system from a broad Gaussian peak to distinct Lorentzian transitions has been interpreted as charge fluctuations shifting the excitonic transition energies. This would cause the broadening of the Lorentzian peaks, hiding them under the same contribution. This hypothesis has been modelled using the equations of a Gaussian stochastic model of spectral diffusion \cite{kubo_ryogo_stochastic_1962, berthelot_unconventional_2006}. This model treats the time dependence of the transition energy as random fluctuations about an average value with Gaussian statistics:
\begin{gather}
   \omega(t) = <\omega> + \delta\omega(t),\\
    <\delta\omega(t)> = 0. 
\end{gather}
These fluctuations in the frequency allow the system to explore a Gaussian distribution of transition frequencies with a variance of $\Sigma = \sqrt{<\delta\omega^2>}$. From here, the correlation time describes the timescale of these random fluctuations:
\begin{gather*}
    \tau_c = \frac{1}{\Delta^2}\int^{\infty}_{0} dt <\delta\omega(t)\delta\omega(0)>.
\end{gather*}
From there, the frequency-dependent intensity spectrum can be obtained by taking the Fourier transform  of the temporal correlation function: 
\begin{gather}
    \sigma(\omega) = \int^{\infty}_{0}e^{-g(t) + it(\omega - \omega_0)}dt,\\
    g(t) = \frac{\Sigma^2\tau_c^2}{\hbar^2}(\frac{t}{\tau} + e^{-\frac{t}{\tau}}- 1),
\end{gather}
where $\omega_0$ represents the frequency that corresponds to each excitonic transition. From the behaviour of the function, two limits can be established. For $\tau_c\Sigma >> \hbar$, the distribution will have a Gaussian lineshape. For $\tau_c\Sigma << \hbar$, a Lorentzian lineshape is found. This behaviour has been observed in our system, but to better understand the equations in detail, the integral has been analytically solved using {\it{Wolfram Mathematica}} to obtain the Fourier transform:
\begin{gather}
    \sigma(\omega) = \tau_ce^\frac{\Sigma^2\tau_c^2}{\hbar^2}(\frac{\Sigma^2\tau_c^2}{\hbar^2})^{-\tau_c(\frac{\Sigma^2\tau_c}{\hbar^2} - i\omega + i\omega_0)}(\Gamma[\tau_c(\frac{\Sigma^2\tau_c}{\hbar^2} - i\omega + i\omega_0)] - \Gamma[\tau_c(\frac{\tau_c\Sigma^2}{\hbar^2} - i\omega + i\omega_0), \frac{\Sigma^2\tau_c^2}{\hbar^2}]),
\end{gather}
where $\Gamma[z]$ and $\Gamma[a, z]$ are the gamma function and incomplete gamma function, respectively. From the experimental data, this model has to be fitted to obtain the values of the respective parameters. Then, the values of the FWHM of the Gaussian spectra (no hBN-capping), the FWHM of the Lorentzian spectra (hBN-capped) and the frequency corresponding to the Lorentzian excitonic transition peaks are used to replicate the experimental data using the model. From there, the first step is to calculate $\Sigma$ for the Gaussian case, denoted here as $\Sigma_G$, which corresponds to the Gaussian variance:
\begin{equation}
    \Sigma_G = \frac{FWHM_G}{2\sqrt{2\ln{2}}}.
\end{equation}
The criteria for the Gaussian limit in the equations allow us to establish an inferior limit for $\tau_c$. With these values and by varying the value of $\Sigma$, the behavioural shift of the curves can be observed, as shown in Figs.~\ref{Fig2}.b and ~\ref{Fig3}.b. Thus, by studying this shift through the lens of spectral diffusion, we can see that the fitted model that we obtain shows good agreement with the experimental data.

However, to understand why this shift occurs and obtain the rest of the parameters, further modelling is needed. In order to reach our goal of being able to measure exciton fine structure through low-defect samples, we tried to study the effects of the active fluctuation centres on the sample through a simple model of N uncorrelated traps, where each trap contributes a Stark shift $\Delta$ of the transition energy \cite{berthelot_unconventional_2006}. In our case, we directly relate $\tau_c$ to the timescale of the processes that affect the spectral diffusion. If we assume that the random fluctuations of the active fluctuation centres follow the same statistics as the random fluctuations considered previously, the previous equations can be directly used. Due to the fact that the value for $\tau_c$ that we obtained through our experimental results is an inferior limit, we compared it to the documented times of molecular reorientation \cite{dahlman_dynamic_2021} and found it to be inferior by two orders of magnitude in the case of our organic spacers. This confirms the fact that we experimentally obtained an inferior limit, but this difference is large enough that we should consider that possible extra effects may not have been taken into account. As this time directly depends on the characteristics of the system, and not on the number of uncorrelated traps, it can be taken as constant, which allows us to extract the value of $\Sigma_L$:
\begin{equation}
    \Sigma_L = \sqrt{\frac{FWHM_L\hbar}{2\tau_c}}.
\end{equation}
From here, and assuming ergodicity, there is a relation between the spectral range modulation amplitude $\Sigma$, the number of active fluctuation centres in the system and the Stark shift caused by a single  centre, $\Delta$ \cite{kubo_ryogo_stochastic_1962, berthelot_unconventional_2006}. By dividing $\Sigma_G$ by $\Sigma_L$, a relation between $\Sigma_G$, $\Sigma_L$ and the active population in each state (uncapped perovskite and hBN-capped perovskite) can be obtained:
\begin{equation}
    \frac{\Sigma_G}{\Sigma_L} = \sqrt{\frac{N_G}{N_L}},
\end{equation}

which allows us to define the following ratio $\mu$:
\begin{equation}
    \mu^{n = 1}_{Exp} = \frac{N_G}{N_L} = (\frac{\Sigma_G}{\Sigma_L})^2.
\end{equation}
The result of this relation is that the active fluctuation centres are reduced by hBN-capping by factors of $\mu^{n = 1}_{Exp} = (3.7 \pm 0.2); (3.7 \pm 0.4); (4.9 \pm 0.4)$ and $(4.4 \pm 0.4)$ for each of the four transitions in the n = 1 case. In the case of n = 2, $\mu^{n = 2}_{Exp} = (4.4 \pm 0.8); (7.3 \pm 1.2); (3.9 \pm 0.6);$ and $(5.1 \pm 0.9)$. As the real value of the characteristic time is unknown, and its inferior limit was what we obtained, our result is just an inferior limit of the active population reduction caused by hBN-capping.

\subsection*{Geometrical matching model}

To further test our hypothesis that van der Waals forces clamp the local movement of the organic spacers of the perovskite, we needed to calculate the geometrical matching between the square lattice of the upper layer of organic spacers and the hexagonal lattice of the bottom layer of hBN that was transferred onto the layer of organic spacers. For this purpose, we used the Python package `pybinding' to generate a model heterostructure, with the top layer twisted by a variable angle $\theta$ in order to account for variations in the positioning of the lattice. This is possible thanks to the symmetries present in square and hexagonal lattices.

As parameters of the model, the sizes of the flakes are relevant, as they limit the calculations of the model on long-range patterns and highly extend calculation times. Additionally, the distances between atoms were taken from \cite{liu_van_2019} and introduced into the model. After creating the model heterostructure, a problem arises, as the exact vertical distance between the layers is not experimentally known; for this reason, we used the average of the possible interatomic distances ($d_{vdW}^{H-N} = 0.312$ nm, $d_{vdW}^{H-B} = 0.320$ nm, $d_{vdW}^{C-N} = 0.352$ nm, $d_{vdW}^{C-B} = 0.359$ nm), which is 0.335 nm. From these distances, the tolerance distance in the lattice plane is defined using the triangle formed by these interatomic distances, which results in a theoretically optimal tolerance distance of $d_T = 0.071$ nm ($0.49d_{hBN}$) between the atoms of the two layers. From there, the distance to every atom from the hexagonal lattice is calculated for each of the organic spacer atoms. For each one of those distances, a comparison with the tolerance limit is done. If the obtained value is under the tolerance distance, a coincidence is counted, and the calculation goes to the next square lattice atom. Once the coincidences are counted for a model, the twisting angle is increased, another model is created and the coincidences are recalculated, with the angle varying from 0 to 60 degrees due to the fact that the symmetry of the hexagonal lattice reduces the angle range needed to cover all possibilities. After the coincidence calculation, the theoretical ratio of clamped molecules was calculated as

\begin{equation}
    \mu_{Theo} = \frac{N_{sq. lat.}}{N_{sq. lat.} - N_{coincidences}}.
\end{equation}

The models were created using different sizes: (10 x 10, 15 x 15) nm, (20 x 20, 30 x 30) nm and (30 x 30, 45 x 45) nm. The smaller size in each pair was used for the perovskite layer and the bigger size was used for the hBN layer. This is due to the possibility of the corners of the flakes not being in contact when the upper layer is twisted, which was an unwanted effect. As previously mentioned, one of the most important factors to take into account was the simulation size, as it was possible that the small models did not take into account particular long-range pattern matching from the lattices. However, the calculation times increased greatly when the size was increased due to the basic algorithm used to calculate the coincidences, and after the results were analysed, a size of (30 x 30, 45 x 45) nm seemed to be the largest size that was needed.

\begin{figure}
    \centering
    \includegraphics[scale = 0.15]{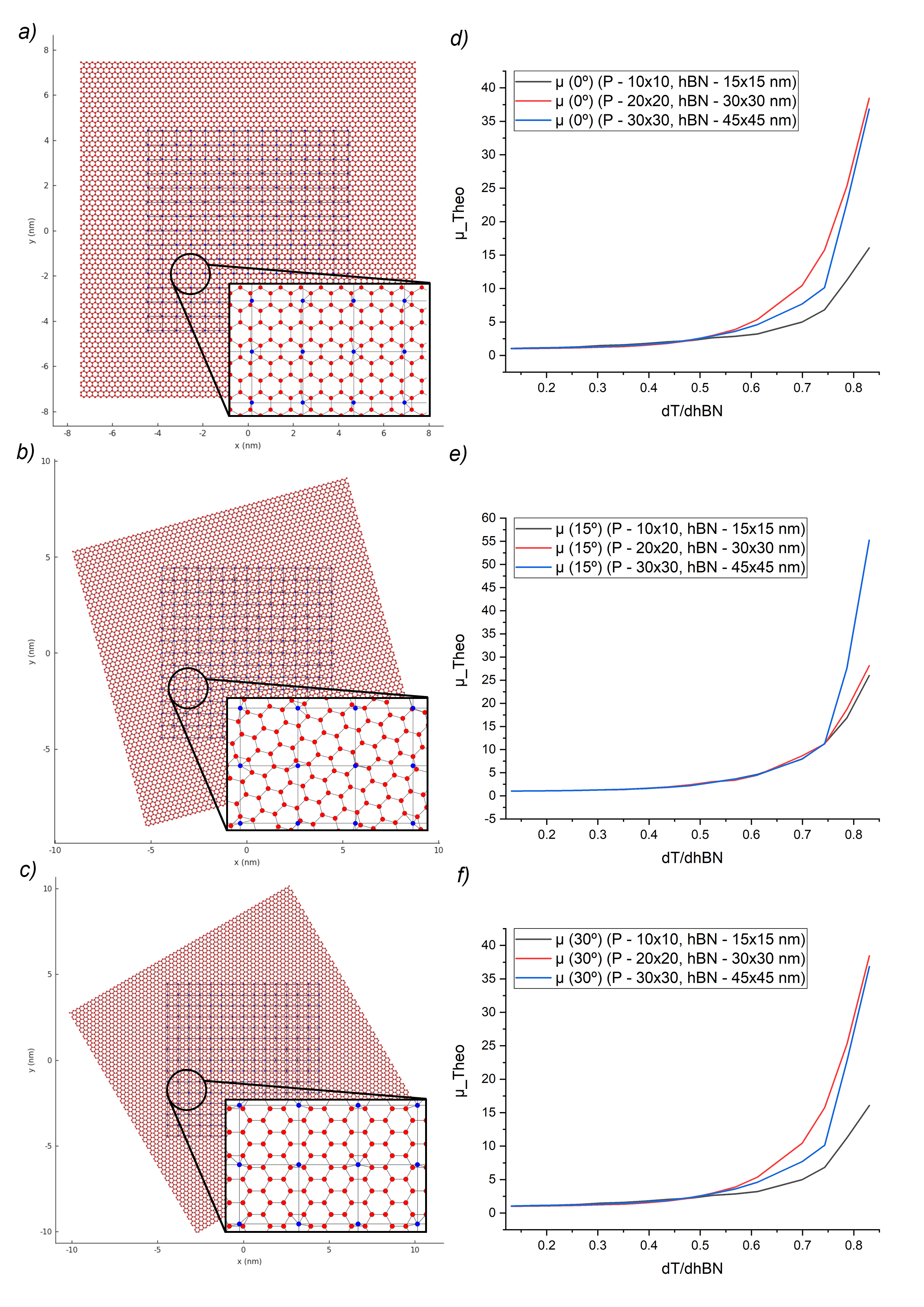}
    \caption{a, b, c) Heterostructure models being twisted by 0, 15 and 30º, with zoomed-in lattices to reflect the twisting angle effect. d, e, f) Calculated values for $\mu_{Theo}$ for 0, 15 and 30º, respectively, for all simulation sizes, as the tolerance distance (in units of $d_{hBN} = 0.145$ nm) increases.}
    \label{S1}
\end{figure}

To account for the effect of the simulation size, we studied the ratio calculations for the different simulation sizes at the calculated angles, as can be seen in Fig. \ref{S1}.d--f. As previously mentioned, the symmetry of the hexagonal lattice reduced the number of angle calculations: a twist angle of 60º results in the same lattice as 0º. Once the symmetries of the square lattice are considered, the angle range needed to cover all lattice configurations is just 30º, as a rotation of 90º of the close-up in Fig. \ref{S1}.c reveals that the resulting lattice is equivalent to our heterostructure with a 0º twisting angle; this can be seen in Fig. \ref{S1}.d, f, which show exactly the same results.

The differences in the behaviour of $\mu_{Theo}$ for different tolerance distances seem negligible when these distances are under half the distance of the hexagonal lattice ($d_{hBN}$) for all angles. However, for 15º, the differences in $\mu_{Theo}$ are negligible until around 0.74 times $d_{hBN}$, where the larger simulation shoots up quickly. Due to the fact that we expect that the larger simulation takes into account these possible long-range patterns, this simulation was used to calculate the ratios discussed in this paper.

\begin{figure}
    \centering
    \includegraphics[scale = 0.15]{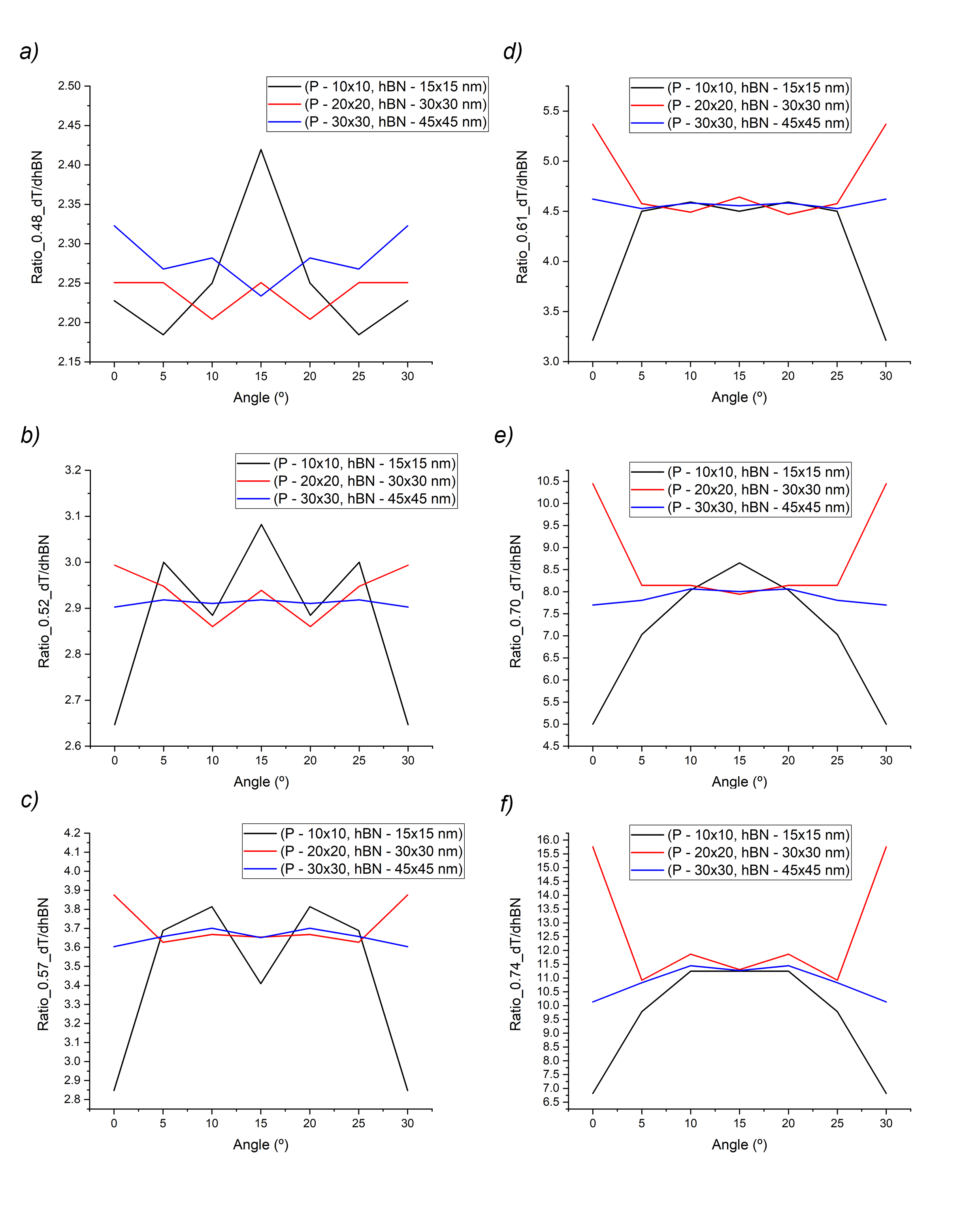}
    \caption{a, b, c, d, e, f) Calculated values for $\mu_{Theo}$ for the entire angle range for six different tolerance distances (0.48, 0.52, 0.57, 0.61, 0.70 and 0.74, in units of $d_{hBN} = 0.145$ nm)}
    \label{S2}
\end{figure}

Finally, to perform our last comparison between the experimental values and our theoretical values, as we do not know the twisting angle present in our heterostructure, we needed to average $\mu_{Theo}$ over all angles to be able to perform any kind of general comparison. However, the problem then is to check if the average is a good approximation to the value for each angle. As Fig. \ref{S2} shows, once the simulation size grows to (30 x 30, 45 x 45) nm and a certain threshold of the tolerance distance is reached, the variation of $\mu_{Theo}$ is extremely small, less than 10\% between the minimal and maximum values, until another threshold is reached: the values again shoot up beyond 0.74 times the tolerance distance.

Thus, it would seem that the average ratio is a good approximation for $\mu_{Theo}$ for the largest simulation size until around 0.72 times the tolerance distance; this is the range in which all of our experimental values were obtained through the photoluminescence measurements lay on, in the theoretical scale of values.

\end{document}